\documentclass[preprint]{aastex}
\usepackage{natbib,psfig,emulateapj5}

\newcommand\nh{\hbox{{$N_{\rm H}$}}}

\newcommand\msun{\hbox{{$M_{\sun}$}}}

\newcommand\rosat{{\sl ROSAT}}
\newcommand\asca{{\sl ASCA}}

\newcommand\chandra{{\sl Chandra}}
\newcommand\xmm{{\sl XMM}}

\newcommand\xspec{{\sc xspec}}

\newcommand\apec{{\sc apec}}

\newcommand\phabs{{\sc phabs}}

\newcommand\ergs{{\rm erg s$^{-1}$}}
\newcommand\kmsmpc{{\rm km s$^{-1}$ Mpc$^{-1}$}}
\newcommand\cmsq{{\rm cm$^{-2}$}}

\newcommand\thot{\hbox{{$T_{\rm h}$}}}
\newcommand\tcool{\hbox{{$T_{\rm c}$}}}
\newcommand\neh{\hbox{{$n_{\rm e}^{\rm h}$}}}
\newcommand\nec{\hbox{{$n_{\rm e}^{\rm c}$}}}

\newcommand\fe{\hbox{{$Z_{\rm Fe}$}}}

\newcommand\si{\hbox{{$Z_{\rm Si}$}}}
\newcommand\su{\hbox{{$Z_{\rm S}$}}}

\newcommand\solar{\hbox{{$Z_{\odot}$}}}

\newcommand\lya{\hbox{{Ly$\alpha$}}}
\newcommand\hea{\hbox{{He$\alpha$}}}

\begin{document} 

\title{An XMM-Newton Observation of NGC 1399 Reveals Two Phases of Hot
Gas and Super-Solar Abundances in the Central Regions}

\author{David A. Buote}

\affil{Department of Physics and Astronomy, University of California
at Irvine, 4129 Frederick Reines Hall,\\ Irvine, CA 92697-4575;
buote@uci.edu}

\slugcomment{Accepted for Publication in The Astrophysical Journal Letters}

\begin{abstract}

We present an initial analysis of a new \xmm\ observation of NGC 1399,
the central elliptical galaxy of the Fornax group. Spectral fitting of
the spatially resolved spectral data of the EPIC MOS and pn CCDs
reveals that a two-temperature model (2T) of the hot gas is favored
over single-phase and cooling flow models within the central $\sim
20$~kpc. The preference for the 2T model applies whether or not the
data are deprojected. The cooler component has a temperature ($\sim
0.9$~keV) similar to the kinetic temperature of the stars while the
hotter component has a temperature ($\sim 1.5$~keV) characteristic of
the virial temperature of a $\sim 10^{13}\msun$ halo. The two-phase
model (and other multitemperature models) removes the ``Fe
Bias'' within $r\la 20$~kpc and gives $\fe/\solar\approx 1.5-2$. At
larger radii the iron abundance decreases until $\fe/\solar\sim 0.5$
for $r\sim 50$~kpc. The Si abundance is super-solar (1.2-1.7 solar)
within the central regions while $\si/\fe\approx 0.8$ over the entire
region studied. These Fe and Si abundances imply that $\approx 80\%$
of the Fe mass within $r\sim 50$~kpc originates from Type Ia
supernovae (SNIa).  This SNIa fraction is similar to that inferred for
the Sun and therefore suggests a stellar initial mass function similar
to the Milky Way.

\end{abstract}

\keywords{X-rays: galaxies: clusters -- galaxies: halos -- galaxies: formation -- cooling flows} 

\section{Introduction}
\label{intro}

NGC 1399, the central galaxy of the Fornax group, is one of the
brightest elliptical galaxies in X-rays and has been the subject of
numerous X-ray studies. Although previous
\rosat\ and \asca\ studies demonstrated that the hot gas within $r\sim
30$~kpc is not isothermal \citep[e.g.,][]{rang95,jone97,buot99a}, the
data could not distinguish between single-phase and multiphase
models. Moreover, we showed that an ``Fe Bias'' for systems like NGC
1399 occurs if an isothermal gas is assumed when in fact the spectrum
consists of multiple temperature components with values near 1 keV
\citep[e.g.,][]{buot00a,buot00c}. We argued that this bias is
primarily responsible for the very sub-solar Fe abundances found in
the central regions of the (X-ray) brightest ellipticals and groups.

The combined spatial and spectral resolution of the \xmm\ CCDs allows
for unprecedented mapping of the temperatures and elemental abundances
of the hot gas in galaxies, groups, and clusters. These capabilities
complement the high spatial resolution ($\sim 1\arcsec$) of \chandra\
which has already provided interesting constraints on the emission
from discrete sources near the center of NGC 1399 \citep{ange01}. The
higher energy resolution, sensitivity, and larger field-of-view of the
\xmm\ EPIC CCDs are better suited for constraining the spatial and
spectral properties of the diffuse hot gas out to radii well past the
optical extent of the galaxy (i.e., out to $r\sim 50$~kpc assuming a
distance of 21~Mpc using the results of \citet{tonr01} for
$H_0=70$~\kmsmpc).

We present initial results for the temperature and metal abundances of
the hot gas obtained from a new \xmm\ observation of NGC 1399. Detailed
discussions of the data reduction, spectral fitting, and results for
the gravitating mass will appear in \citet{buot02d} and \citet{lewi02b}.

\section{Spatial-Spectral Analysis}
\label{analysis}

NGC 1399 was observed with the EPIC pn and MOS CCD cameras for
approximately 25~ks and 30~ks respectively during AO-1 as part of the
\xmm\ Guest Observer program. We generated calibrated events lists
for the data using the standard SAS v5.3.0 software. Our data
preparation closely follows standard procedures and will be described
fully in a follow-up paper \citep{buot02d}.

We extracted spectra in concentric circular annuli located at the X-ray
centroid of each detector. The widths of the annuli were the same for
each detector and were defined such that the width of each annulus
contained approximately 6000 background-subtracted counts in the
0.3-5~keV band in the MOS1, and the minimum width was set to
$1\arcmin$ for PSF considerations. Obvious point sources were masked
out before the extraction.

In each annulus we performed a simultaneous fit to the MOS1, MOS2,
and pn spectral data over 0.3-5~keV. Our baseline model (1T) consists
of a single thermal plasma component using the \apec\ code modified by
foreground Galactic absorption ($\nh=1.3\times 10^{20}$~\cmsq) using
the \phabs\ model in \xspec\ v11.1.0v. The free parameters are all
associated with the plasma component: temperature ($T$),
normalization, and Fe, O, Mg, and Si abundances -- all other elements
tied to Fe in their solar ratios except for S which is tied to Si in
their solar ratio. (The normalizations for the MOS1 and MOS2 are tied
together while that of the pn is varied separately because of residual
calibration and background differences.) We obtained results for the
1T model fitted directly to the data projected on the sky (i.e., 2D
model) and also fitted to the deprojected data (i.e., 3D model)
\citep[e.g.,][]{buot00c}.

\begin{figure*}[t]
\parbox{0.32\textwidth}{
\centerline{\psfig{figure=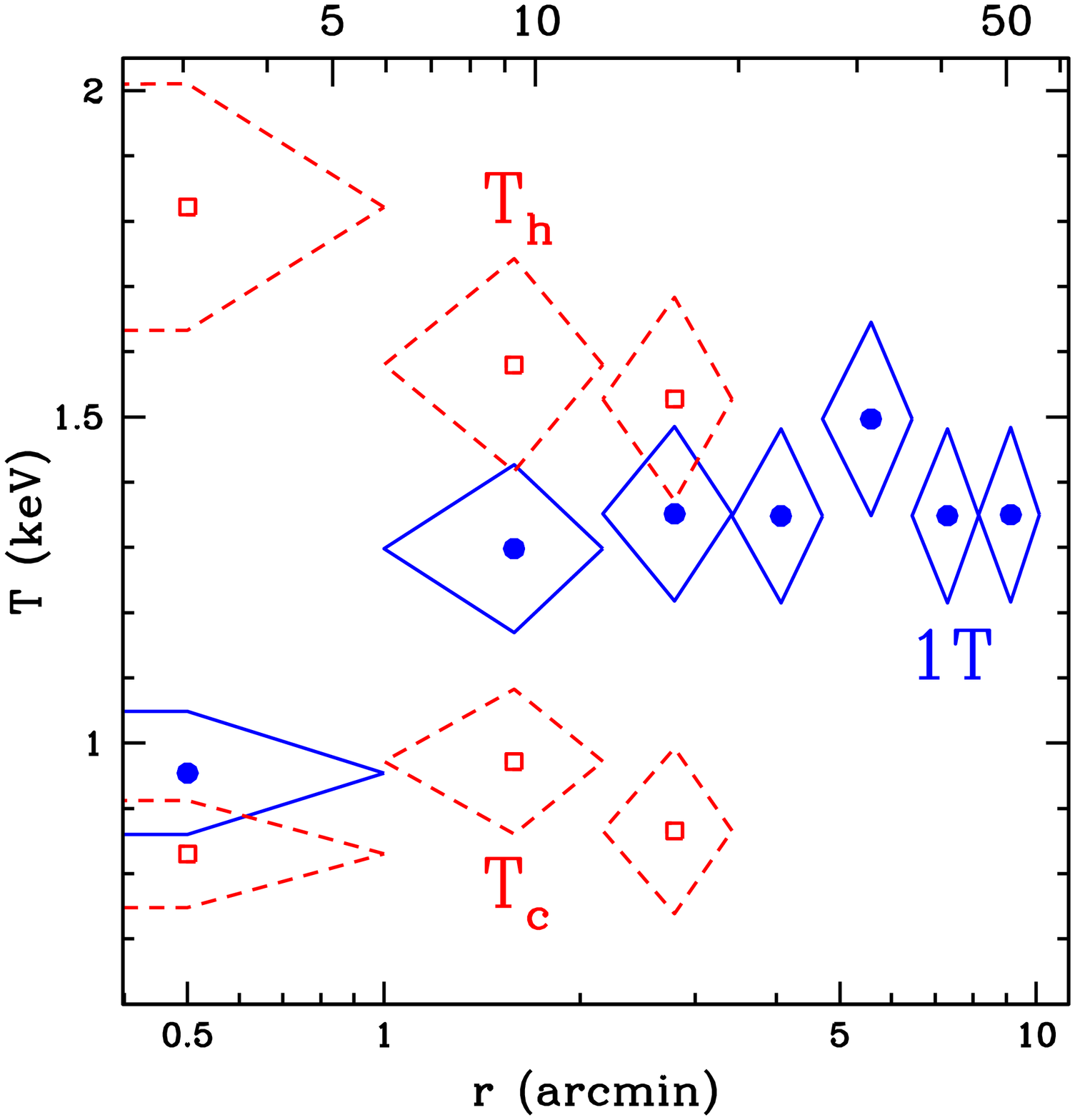,height=0.24\textheight}}
}
\parbox{0.32\textwidth}{
\centerline{\psfig{figure=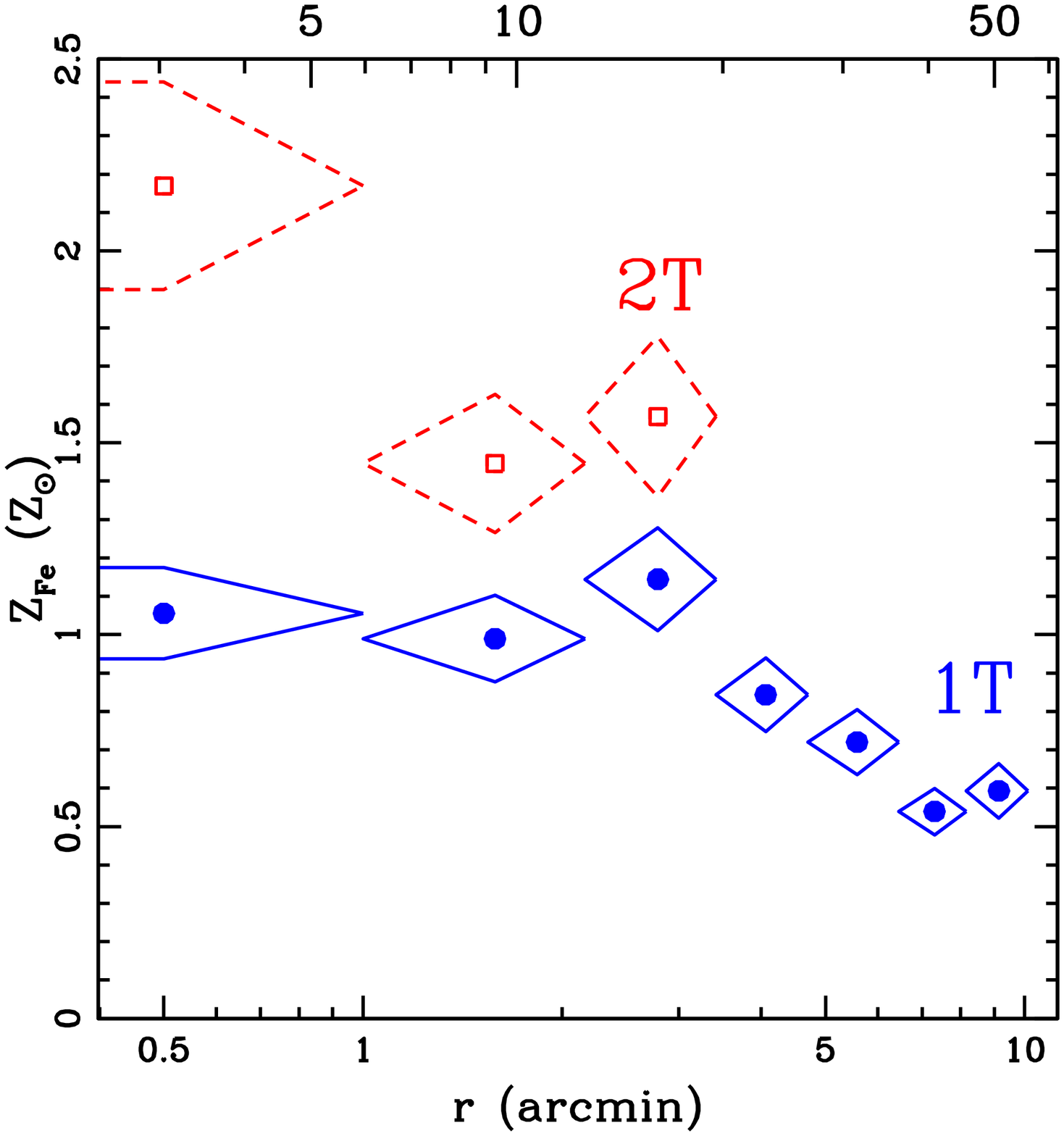,height=0.24\textheight}}
}
\parbox{0.32\textwidth}{
\centerline{\psfig{figure=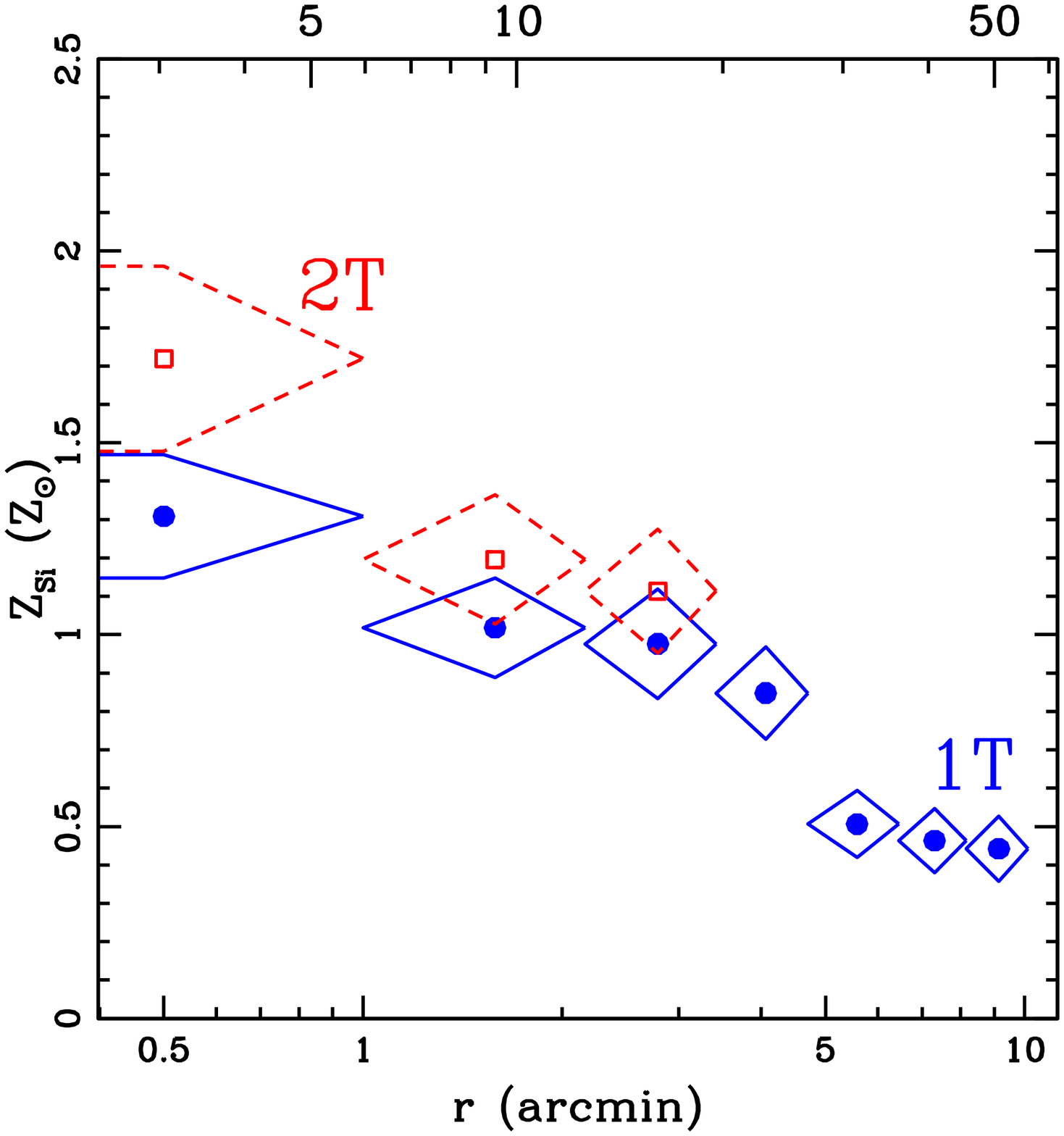,height=0.24\textheight}}
}
\caption{\label{fig} ({\sl Left panel}) Radial temperature profiles
(units -- bottom: arcminutes, top: kpc) and $1\sigma$ errors for 1T
and 2T models (both 2D). ({\sl Middle panel}) The iron abundance
profiles and ({\sl Right panel}) the silicon abundance
profiles of these models with $1\sigma$ errors.}
\end{figure*} 

\begin{figure*}[t]
\parbox{0.49\textwidth}{
\centerline{\psfig{figure=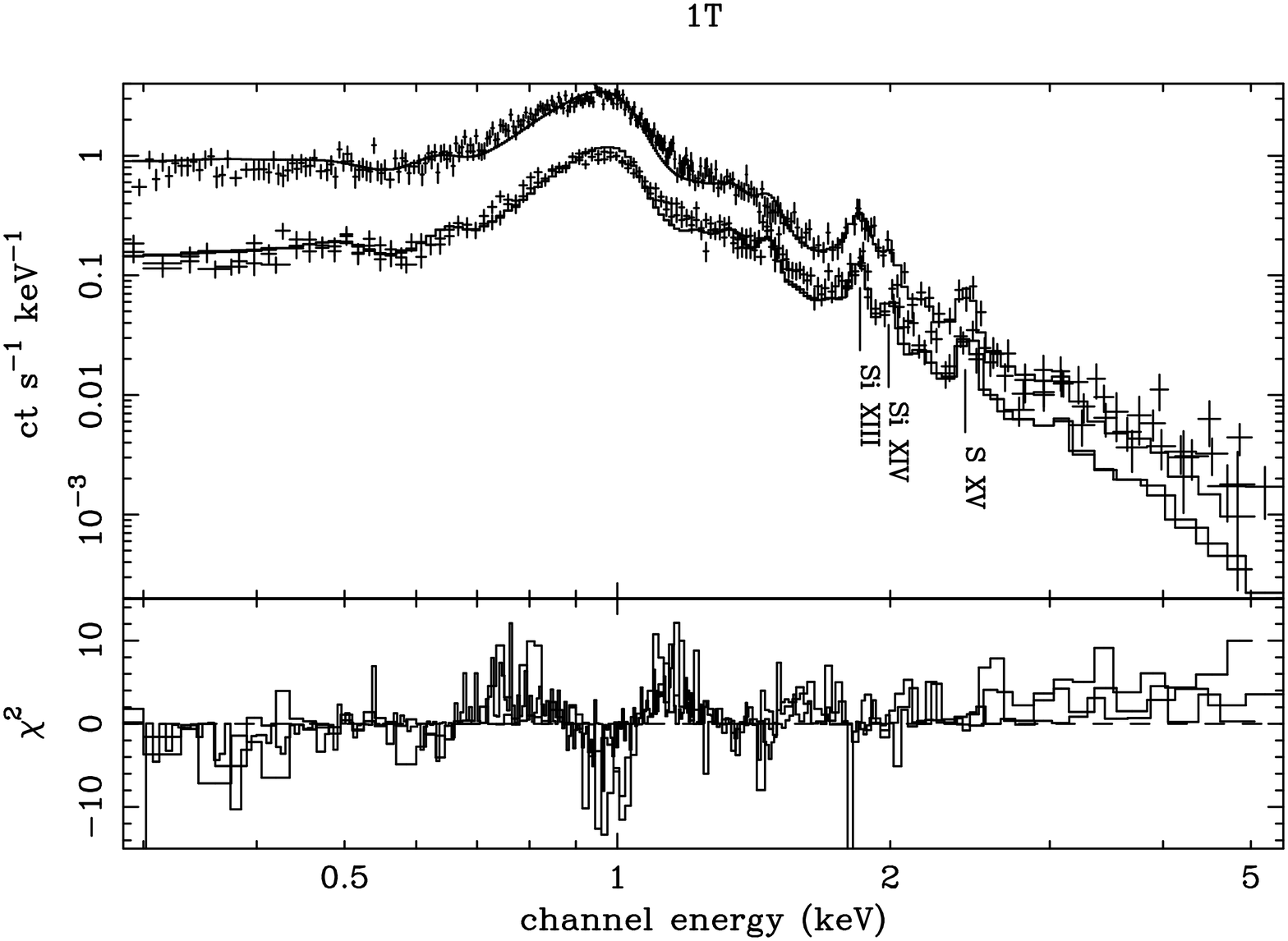,height=0.24\textheight}}
}
\parbox{0.49\textwidth}{
\centerline{\psfig{figure=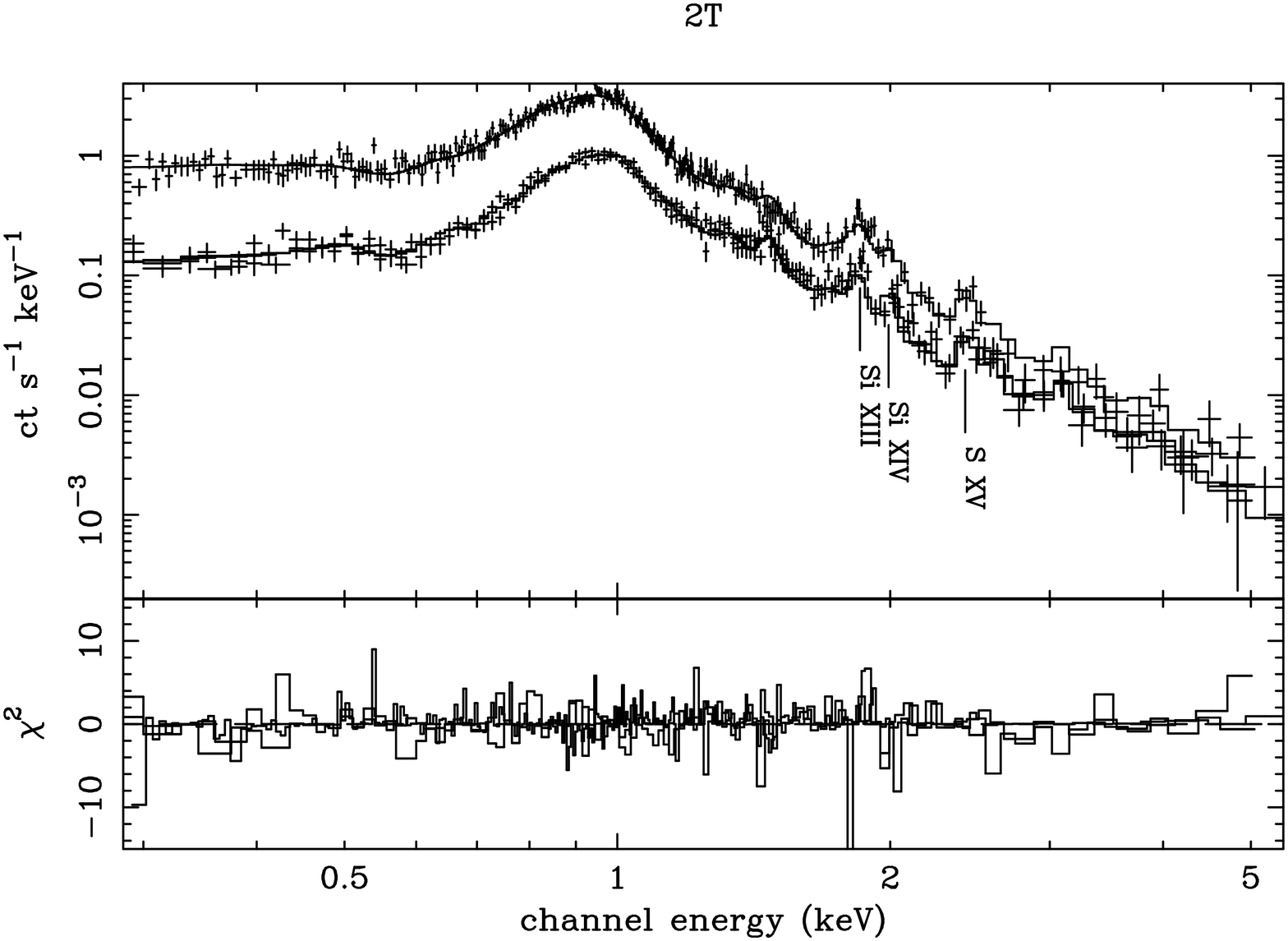,height=0.24\textheight}}
}
\caption{\label{fig.spec} MOS and pn spectra for annulus \#1 fitted
with ({\sl Left panel}) a 1T (2D) model and ({\sl Right
panel}) a 2T (2D) model.}
\end{figure*}

\begin{table*}[t] \footnotesize
\caption{Quality of Spectral Fits
\label{tab}}
\begin{center} \vskip -0.4cm
\begin{tabular}{ccccccc} \tableline\tableline\\[-7pt]
& $R_{\rm in}$ & $R_{\rm out}$ & \multicolumn{2}{c}{1T} &
\multicolumn{2}{c}{2T}\\ 
Annulus & (arcmin) & (arcmin) & $\chi^2$ & dof & $\chi^2$ & dof\\ 
\tableline \\[-7pt]
1 & 0.0   &  1.0  &  1061.3   &	487 &	   558.2  &     484\\ 
2 & 1.0   &  2.2  &   700.1   &	520 &	   625.5  &     517\\ 
3 & 2.2   &  3.4  &   691.8   &	593 &	   657.8  &     590\\ 
4 & 3.4   &  4.7  &   824.4   &	645 &	   &     \\ 
5 & 4.7   &  6.5  &   925.3   &	776 &	   &     \\ 
6 & 6.5   &  8.2  &   821.1   &	826 &	   &     \\ 
7 & 8.2   &  10.1 &  1042.8   &	907 &	   &     \\ 
\tableline \\[-1.0cm]
\end{tabular}
\end{center}
\end{table*}
\subsection{Two-Phase Temperature and Density Structure}
\label{temp}

The temperature profile for the 1T (2D) model is displayed in Figure
\ref{fig}. The temperature rises from $T\sim 0.9$~keV at the center to
$T\sim 1.4$~keV at large radii consistent with previous 2D \rosat\
determinations \citep[e.g.,][]{rang95,jone97,buot99a}. This model is a
good fit in the outer annuli but the fit degrades at progressively
smaller radii (see Table \ref{tab}). The worst fit is in the central
annulus and, as shown in Figure \ref{fig.spec}, yields fit residuals
near 1~keV that are fully characteristic of those obtained when trying
to force a single-temperature model to fit a multi-temperature
spectrum \citep[e.g.,][]{buot98c,buot99a,buot00a}. The deviations
above 2~keV also suggest the presence of another (higher) temperature
component.

Multitemperature spectra in the inner regions, especially the central
bin, can arise simply from the projection of hotter gas at larger
radius. In this case fitting the 1T model to the deprojected spectra
should substantially improve the fits in the central regions. However,
we find that this 1T (3D) model does not improve the fits much except
at the center, where noticeable -- though relatively modest --
improvement ($\Delta\chi^2= 146$) is achieved.

We have examined a variety of multitemperature models (2D and 3D) and
have found that simple two-temperature models (2T) provide the best
fits to the central annuli (Table \ref{tab}). It is also found that
the $\chi^2$ values are very similar and the derived parameters
consistent within their $1\sigma$ errors for both 2D and 3D versions
of the 2T models. And since the parameters of the 3D models are not as
well constrained and require regularization of the temperatures and
abundances \citep[cf.][]{buot00c}, in the interest of brevity we will
emphasize in this letter the results for the 2D models.  Detailed
results for the 3D versions of 1T, 2T, and other models will appear in
\citet{buot02d}.

With respect to the 1T (2D) model, not only does the 2T (2D) model
improve the fit substantially within the central bin
($\Delta\chi^2=503$), but it also significantly improves the fits in
annuli \#2 ($\Delta\chi^2=75$) and \#3 ($\Delta\chi^2=34$). The
dramatic improvement in the central bin is seen in Figure
\ref{fig.spec} where the large residuals in the Fe L region near 1 keV
have been mostly eliminated. In 3D the improvement of 2T over 1T is
$\Delta\chi^2=363$ in the central bin. The other multitemperature
models we examined only improved fits to the deprojected spectra in
the central bins; e.g., a (constant pressure) cooling flow model
consisting of ambient and cooling gas provides a modest improvement to
the fit in the central bin ($\Delta\chi^2=99$) over the 1T (3D) model.

In Figure \ref{fig} are shown the temperatures obtained for the 2T
(2D) model. The cool component is consistent with an isothermal radial
profile while the hot component appears to rise slightly at the
center. Within the $1\sigma$ errors this central rise for the hot
component is not especially significant, but if real the slight excess
over the isothermal extrapolation from the adjacent annuli probably
represents the contribution from hard emission from discrete sources
unresolved by \xmm\ but resolved in the \chandra\ image of NGC 1399
\citep{ange01}. We mention that adding an additional bremsstrahlung
component to account for the effect of these unresolved sources (i.e.,
2T+BREM) in the central bin does improve the fit slightly and gives a
reasonable best-fitting temperature, $T\approx 5$~keV. Adding this
component to the cooling flow model instead gives $T\approx 2.7$~keV
indicating an attempt to account for both $\sim 1$~keV hot gas and
$\sim 5$~keV unresolved sources.

{\it We believe that in addition to the $\chi^2$ values the derived
parameter values also argue in favor of the 2T model.}  The
temperature of the hot component, $\thot\sim 1.5$~keV, is consistent
with the virial temperature of a surrounding group of mass $\approx
10^{13}\msun$, whereas the temperature of the cool component, $\tcool
\sim 0.9$~keV, is similar to the kinetic temperature of the stars
($T\sim 0.6$~keV) with velocity dispersion, $\sigma=310$~km~s$^{-1}$
\citep[e.g.,][]{jone97}. The volume-averaged electron density of the
cool gas (\nec) declines with radius more rapidly than that of the hot
component (\neh): for the best-fitting models the ratio $\nec/\neh$ is
1.0, 0.62, and 0.27 respectively for shells \#1, \#2, and \#3. (Here
we have used the 3D model so that the quoted densities refer to the 3D
spherical shells bounded by the same radii as the annuli listed in
Table \ref{tab}.) The cool component is only detected out to annulus
\#3 (i.e., for $r\la 20$~kpc) and therefore must contribute even less
at larger radius. The radius where the cool (galaxy) component gives
way to the hot (group) component is similar to that inferred from the
recent analysis of the \rosat\ surface brightness profile of NGC 1399
by \citet{paol02b}.

Interior to shell \#3 we compute a gravitating mass of $\approx
10^{12}\msun$ from each component consistent with previous X-ray
\citep[e.g.,][]{rang95,jone97,paol02b} and optical
\citep[e.g.,][]{sagl00} determinations. The consistency of the masses
determined from each phase, and the agreement with optical
determinations, indicates that each phase is in approximate
hydrostatic equilibrium tracing the same gravitational potential.

Although hydrostatic equilibrium implies subsonic gas motions for each
phase, the phases are not in pressure equilibrium.  The ratio of the
gas pressure in the hot phase to that of the cool phase is $\approx
2.4$ in shell \#1 which increases to $\approx 6.9$ in shell \#3. The
lack of pressure equilibrium implies that the low-pressure cool phase
must be continuously replenished. This requirement is consistent with
the scenario where the cool phase continuously receives gas from
stellar ejecta and loses gas by eventually mixing with the extended
hot phase. Detailed physical models for the two phases (including,
e.g., appropriate filling factors) will be considered in a future
study.

\subsection{Fe Abundance}
\label{fe}

We take the solar abundances in \xspec\ to be those given by the
\citet{angr} table except for Fe where we use the correct
``meteoritic'' value, $\rm Fe/H=3.24\times 10^{-5}$ by number, as
suggested by \citet{ishi97} and \citet{mcwi97}. This value is 1.44
times smaller than the photospheric value which is usually employed in
X-ray spectral fitting. Consequently, we shall account for this factor
when comparing our results to other studies.

The Fe abundances (\fe) obtained for the 1T and 2T models (both 2D)
are displayed in Figure \ref{fig}. (The values of \fe\ obtained from
2D and 3D 2T models are consistent within 1-2~$\sigma$ errors.) At the
largest radii the abundance is sub-solar, $\fe/\solar\approx
0.5-0.6$. It increases for smaller $r$ so that within $r\sim 20$~kpc,
$\fe/\solar\approx 1$, for the 1T model similar to previous results
for 1T (2D) models fitted to the \rosat\ data
\citep[e.g.,][]{jone97,buot99a}.

Compared to the 1T model, the 2T model gives \fe\ that is 40\%-50\%
larger in annuli \#2 and \#3 and twice as large in annulus \#1.  (A
cooling flow model, when fitted to the deprojected spectra, produces
similarly larger values.) This substantial increase in the value of
\fe\ along with the elimination of the characteristic $\sim 1$~keV
residual pattern (Figure \ref{fig.spec}) is a manifestation of the
``Fe Bias'' we have discussed previously for NGC 1399 \citep{buot99a}
and related systems \citep[e.g.,][]{buot00a,buot00c}.  We note that
the abundances in each temperature component are tied together in the
fits since they cannot be distinguished when fitted separately. Also,
the values of
\fe\ deduced in the central bins for both 1T and 2T models differ
slightly if intrinsic absorption is considered (which is not clearly
required by the data as seen in Figure \ref{fig.spec}), but the relative
differences between the 1T and 2T models and the general finding of
super-solar values are preserved.

\subsection{Si and S Abundances}
\label{si}

Next to the broad feature of Fe L lines near 1 keV, the most notable
spectral lines in the EPIC spectra of NGC 1399 (Figure \ref{fig.spec})
are the K$\alpha$ lines of Si and S; i.e., \ion{Si}{13} \hea\
(1.85~keV), \ion{Si}{14} \lya\ (2.0~keV), and \ion{S}{15} \hea\
(2.45~keV). We find that \si\ is generally better constrained than
\su, but when these abundances are fitted separately they inevitably
yield values that are consistent within their 1-2~$\sigma$ errors. To
reduce the number of free parameters in our fits we tie \si\ and \su\
together in their solar ratio. Henceforth, we shall refer only to the
Si abundance.

The Si abundance is plotted in Figure \ref{fig} for both 1T and 2T
models. For the 1T model $\si/\solar\approx 0.5$ for $r\ga 30$~kpc,
rises to values consistent with solar for $r\sim 5-20$~kpc, and is
slightly larger than solar in the central bin. For the 1T model
$\si/\fe\la 1$ at large radii, $\si/\fe\sim 1$ for $r\sim 5-20$~kpc,
and $\si/\fe\ga 1$ in the central bin.

The 2T model gives slightly larger values, $\si/\solar\sim 1.1-1.2$
for $r\sim 5-20$~kpc (which are consistent with the 1T values within
the $1\sigma$ errors). But in the central bin the 2T value is $\approx
30\%$ larger; i.e., about a $2\sigma$ discrepancy.  In contast to the
1T case, for the 2T model, $\si/\fe\approx 0.8$ for $r\la
20$~kpc. (The overestimate of $\si/\fe$ using the 1T model is a ``Si
Bias'' related to the Fe bias we discussed previously for such systems
\citep{buot00a}.)

(Unlike Fe and Si, the O and Mg abundances are quite sensitive to
intrinsic absorption. We will discuss models with intrinsic absorption
in \citet{buot02d}.)

\section{Conclusions}
\label{conc}

We have performed a spatially resolved spectral analysis of the \xmm\
EPIC data of NGC 1399 which suggests the presence of two dominant
phases of hot gas within the central $\sim 20$~kpc. This two-phase
description is consistent with 2T models we fitted previously to the
\asca\ data  \citep{buot99a}. However, the \asca\ data lacked the spatial
resolution of \xmm\ and could not distinguish between a radially
varying single-phase medium, a two-phase medium, or a cooling flow
within $r\sim 30$~kpc. These results for NGC 1399 are supported by
previous analyses of its X-ray surface brightness profile which argues
for separate galaxy and group components
\citep[e.g.,][]{ikeb96,paol02b} similar to other centrally E-dominated
groups \citep{mulc98}.

Since the temperature structure of NGC 1399 obtained by \asca\ is
characteristic of other central E-dominated galaxy groups
\citep{buot99a,buot00a,buot99b,alle00}, the two-phase structure
revealed by \xmm\ for NGC 1399 likely applies to these other
systems. Moreover, since the centers of several clusters show similar
multitemperature structure with \asca\ \citep[e.g.,][]{buot99b,alle01}
and similar multi-component surface brightness structure
\citep[see][and references therein]{etto00}, we might expect two-phase
structures in the centers of these systems. A recent \xmm\ analysis of
M87 is consistent with this suggestion \citep{mole01b}.

The two-phase model provides a physical explanation for the gas
density and temperature profiles seen in ``cooling flows''. The hotter
component is associated with the ambient group or cluster gas while
the cooler component is associated with the stellar ejecta of the
dominant central galaxy. Gasdynamical models of groups and clusters
constructed by \citet{brig99b,brig02a} demonstrate that relatively
cool ejecta from stellar mass loss combined with the hotter ambient
gas can reproduce observed single-phase temperature profiles in groups
and clusters. But our results suggest that these two phases are not
fully mixed.

Understanding the details of the interaction of these phases may shed
light on the mystery of cooling flows. Using the results of
\citet{math89}, the rate of energy input by stellar ejecta is,
$\approx\alpha_{\star}M_{\star}\sigma^2$, where $\alpha_{\star}\approx
5\times 10^{-20}$~s$^{-1}$ is the specific stellar mass-loss rate. If
within $r\sim 5$~kpc we take $M_{\star}\approx 2\times 10^{11}\msun$
for the stellar mass and $\sigma=310$ km s$^{-1}$ \citep{jone97}, we
infer the rate of energy input into the central $\sim 5$~kpc of NGC
1399 from stellar ejecta to be $\approx 2\times 10^{40}$~\ergs. This
crude estimate is only a factor of two less than the X-ray luminosity
of the hot component (and a factor of four less than the cool
component). If the energy input from stellar ejecta does not
completely suppress mass drop-out, it may at least delay the episodes
of AGN feedback. Problems with such heating models are discussed by
\citet{brig02b}.

The two-phase model (as well as other multitemperature models) removes
the ``Fe Bias'' within the central regions ($r\la 20$~kpc) and gives
$\fe/\solar\approx 1.5-2$. These super-solar values exceed the stellar
values in typical elliptical galaxies \citep[e.g.,][]{trag00a} and
therefore allow for significant enrichment of the hot gas from Type Ia
supernovae (SNIa) \citep[cf.][]{arim97}. For example, using the
results of \citet{gibs97} and our measured value of $\si/\fe \approx
0.8$ solar, we infer that SNIa have contributed $\approx 80\%$ of the
iron mass within $r\sim 50$~Kpc in NGC 1399. This SNIa fraction is
similar to that inferred for the Sun and therefore suggests a stellar
initial mass function similar to the Milky Way as advocated by Renzini
and others \citep[e.g.,][]{renz93,renz97,wyse97}. We close by noting
that although correcting for the ``Fe Bias'' partially removes the
``Iron Discrepancy'' noted by \citet{arim97}, chemical models of
elliptical galaxies without cooling flows predict central iron
abundances even larger than we have measured for NGC 1399
\citep[e.g.,][]{brig99a}.

\acknowledgements It is a pleasure to thank F. Brighenti,
W. Mathews, and the anonymous referee for comments on the manuscript.

\bibliographystyle{apj}

\end{document}